\documentclass[12pt]{article}
\usepackage{draft} 
\usepackage[weather, alpine]{ifsym}

\usepackage{fontawesome}
\usepackage{hyperref}
\usepackage{graphicx,color,subfig}
\usepackage{cite}
\usepackage{mciteplus}
\usepackage{skak}
\usepackage{empheq}
\usepackage{tikz}
\usepackage{bbm}
\usepackage{url}

\usepackage{cite}
\usepackage{mciteplus}
\usepackage{array}
\usepackage{bbm}
\usepackage{simpler-wick}
\usepackage{amsmath}
\usepackage{amsfonts}
\usepackage{amsmath}
\usepackage{graphicx}
\usepackage{float}
\usepackage{amssymb}
\usepackage{hyperref}
\usepackage{mathtools}
\usepackage{cleveref}
\usepackage{extarrows}
\usepackage{relsize}
\usepackage{bbold}
\usepackage{xcolor}
\usepackage{amsthm}

\newcommand{\pd}[1]{\partial}
\newcommand{\tx}[1]{\tilde{x}}
\newcommand{\beq}{\begin{equation}}
\newcommand{\eeq}{\end{equation}}

\newcommand{\cO}{\mathcal{O}}

\newcommand{\cH}{\mathcal{H}}

\newcommand{\corr}[1]{\langle #1 \rangle_\beta}

\def\ie{\begin{equation}\begin{aligned}}
\def\fe{\end{aligned}\end{equation}}

\def\XXint#1#2#3{{\setbox0=\hbox{$#1{#2#3}{\int}$}
		\vcenter{\hbox{$#2#3$}}\kern-.5\wd0}}

\begin{document}

\unitlength = .8mm

\begin{titlepage}
	
	\begin{center}
		
		\hfill \\
		\hfill \\
		\vskip 1cm

		\title{Thermal Bootstrap of Matrix Quantum Mechanics}
		
		\author{Minjae Cho$^{\text{\Snow}\text{\Lightning}}$, Barak Gabai${}^{\text{\IceMountain}}$, Joshua Sandor${}^{\text{\FilledWeakRainCloud}}$, Xi Yin${}^{\text{\FilledWeakRainCloud}}$}
		
		\address{
			$^{\text{\Snow}}$Kadanoff Center for Theoretical Physics \& Enrico Fermi Institute, University of Chicago,\\ Chicago, IL 60637,
			USA
			\\
			$^{\text{\Lightning}}$Princeton Center for Theoretical Science, Princeton University, \\ Princeton, NJ 08544,
			USA
			\\
			$^{\text{\IceMountain}}$Laboratory for Theoretical Fundamental Physics, EPFL, Lausanne, Switzerland
			\\
			${}^{\text{\FilledWeakRainCloud}}$Jefferson Physical Laboratory, Harvard University, \\
			Cambridge, MA 02138 USA
			
		}
		
		\email{cho7@uchicago.edu, barak.gabai@epfl.ch,  jsandor@fas.harvard.edu,  xiyin@fas.harvard.edu}
		
	\end{center}

	\abstract{ We implement a bootstrap method that combines stationary state conditions, thermal inequalities, and semidefinite relaxations of matrix logarithm in the ungauged one-matrix quantum mechanics, at finite rank $N$ as well as in the large $N$ limit, and determine finite temperature observables that interpolate between available analytic results in the low and high temperature limits respectively. We also obtain bootstrap bounds on thermal phase transition as well as preliminary results in the ungauged two-matrix quantum mechanics.
	}
	
	\vfill
	
\end{titlepage}

\eject

\begingroup
\hypersetup{linkcolor=black}

\tableofcontents

\endgroup

\section{Introduction}

The bootstrap approach to quantum and statistical models by combining analytic constraints with convex optimization has been developed in quantum chemistry \cite{PhysRevA.57.4219,10.1063/1.1360199}, conformal field theory \cite{Rattazzi:2008pe}, classical dynamics \cite{Fantuzzi_2016}, and more recently applied to solving random matrix models \cite{Lin:2020mme, Kazakov:2021lel}, matrix quantum mechanics (MQM) \cite{Han:2020bkb, Lin:2023owt}, and lattice quantum field theories \cite{Anderson:2016rcw, Anderson:2018xuq, Kazakov:2022xuh, Cho:2022lcj, Cho:2023ulr,Kazakov:2024ool,Li:2024wrd}. Not bound by the limitations of conventional analytic methods based on perturbation theory or symmetries, nor conventional numerical methods based on Monte Carlo sampling, the bootstrap approach offers a new toolkit for grasping the strongly coupled and chaotic dynamics underpinning holographic dualities and quantum gravity. Thus far, attempts of bootstrapping holographic MQM such as the BFSS model \cite{Polchinski:1999br,Lin:2023owt} has been sharply limited to planar observables in the microcanonical ensemble, e.g. the expectation value of a single-trace operator in the ground state, that by themselves do not unambiguously capture dual gravitational observables.\footnote{For the singlet sector of Hermitian multi-matrix QM, an alternative loop-space approach has been developed in \cite{Koch:2021yeb} to solve observables at large $N$.} It would be of substantial interest to bootstrap, for instance, the entropy of the BFSS model in the canonical ensemble at finite temperature that captures black hole thermodynamics in the holographic dual, and compare with the results obtained from Monte Carlo methods \cite{Hanada:2007ti,Catterall:2007fp,Anagnostopoulos:2007fw,Catterall:2008yz,Hanada:2008gy,Hanada:2008ez,Catterall:2009xn,Hanada:2009ne,Hanada:2013rga}.

In this paper, we take the initial steps toward bootstrapping the thermodynamics of MQM, focusing primarily on the {\it ungauged} one-matrix QM described by the Hamiltonian
\ie\label{1matH}
H = {\rm Tr} \Big[ {1\over 2} P^2 + V(X) \Big],
\fe
where the canonical coordinate $X$ and its conjugate momentum $P$ are Hermitian traceless $N\times N$ matrices. 
In contrast to the gauged version of the MQM (a.k.a. the ``singlet sector'') which admits a well-known reformulation in terms of non-interacting fermions, the ungauged MQM includes ``non-singlet sectors'' that admit a dual description in terms of ``long strings'' \cite{Maldacena:2005hi, Balthazar:2018qdv} and is {\it not} known to be analytically solvable.

Denote by $\corr{\cO}$ the thermal expectation value of an operator ${\cal O}$ at inverse temperature $\B$. Given the Hamiltonian $H$, the stationary state conditions
\ie\label{SDeq}
-i \corr{[H, \widetilde\cO]} = 0
\fe 
for any set of Hermitian operators $\widetilde{\cal O}$ provide a set of linear constraints on the thermal expectation values of Hermitian operators with real coefficients. Furthermore, given any set of operators ${\cal O}_i$, the matrix
\ie\label{adefs}
A_{ij}:= \corr{{\cal O}_i^\dagger {\cal O}_j}
\fe
is positive semidefinite, namely
\ie\label{sqrPositivity}
A\succcurlyeq 0 .
\fe
The conditions (\ref{SDeq}) and (\ref{sqrPositivity}) are analogous to those considered in \cite{Han:2020bkb}. As functional relations on the thermal expectation values, however, they do not depend explicitly on the temperature and clearly cannot by themselves constrain the temperature dependence of any thermal observable.

The key new recipe is a set of convex logarithmic inequalities concerning thermal expectation values, constructed as follows. In addition to (\ref{adefs}), one defines the matrices
\ie\label{ABCmats}
B_{ij} := \corr{\cO_j\cO_i^\dagger},~~~~
C_{ij} := \corr{\cO_i^\dagger [H, \cO_j]}.
\fe
It has been established in \cite{Araki:1977px,Sewell1977,Fawzi:2023fpg} that the semidefinite condition
\ie\label{kmseqn}
\beta C - A^{1/2}\log\Big( A^{1/2} B^{-1} A^{1/2}\Big)A^{1/2} \succcurlyeq 0
\fe
holds, and is in fact equivalent to the Kubo-Martin-Schwinger (KMS) condition \cite{1957JPSJ...12..570K,1959PhRv..115.1342M,1967CMaPh...5..215H} on a thermal state.\footnote{See also Theorem 5.3.15 in \cite{Bratteli:1996xq}. 
} 
Now (\ref{SDeq}) for any set of Hermitian operators $\widetilde{\cal O}$, together with (\ref{sqrPositivity}) and (\ref{kmseqn}) on a finite set of basis operators ${\cal O}_i$, define within the linear space of {\it trial} expectation values a convex subset to which the expectation values in a thermal state must belong.
Given a Hermitian operator ${\cal O}$, minimizing (maximizing) $\corr{\cO}$ on the space of trial expectation values subject to these convex conditions then produces a rigorous lower (upper) bound on the admissible values of $\corr{\cO}$.\footnote{Moreover, it was shown in \cite{Fawzi:2023fpg} that for translation-invariant lattice Hamiltonians with a finite-dimensional local Hilbert space and finite-range interactions, the said bound converges to the physical value of $\corr{\cO}$ as the set of basis operators ${\cal O}_i$ is increased systematically.} 
We refer to this procedure as {\it KMS optimization}.

Crucially, despite that the LHS of (\ref{kmseqn}) is not linear with respect to the thermal expectation values, it was shown in \cite{Fawzi_2018, Fawzi:2023fpg} that the KMS optimization can be reformulated as a standard semidefinite programming (SDP) problem using semidefinite relaxation of the matrix logarithm. We review the relevant algorithm and demonstrate its implementation in the simple example of an aharmonic oscillator at finite temperature in section \ref{sec:TB}.

Applying the same strategy to the ungauged MQM (\ref{1matH}), we consider in section \ref{sec:themalmqm} a truncated basis ${\cal B}$ of adjoint-valued operators ${\cal O}_i$ of the form $\{1, X, P, X^2, XP, PX,\cdots\}$, and construct the matrices $A, B, C$ of (\ref{adefs}), (\ref{ABCmats}) whose entries are the thermal expectation values of operators up to maximal word length $L$. We then employ the semidefinite relaxation of \cite{Fawzi_2018, Fawzi:2023fpg} to set up the SDP problem for the energy expectation value 
\ie
E(\B) \equiv \corr{H},
\fe
at finite $N$ as well as in the planar limit. The upper and lower SDP bounds on $E(\B)$ are seen to converge with increasing maximal word length $L$. For finite $N\leq 10$, we have obtained the bootstrap bounds with $L=8$. In particular, the $N=2$ results are seen to agree with that of Hamiltonian truncation. In the planar limit at infinite $N$, we have obtained the bootstrap bounds with maximal word length $L=10$, where the upper and lower bounds on $E(\B)$ differ by a factor of order $10^{-3}$ at intermediate temperatures. 

The planar bootstrap results are further compared in section \ref{sec:UMQM} to the analytic high temperature expansion (based on perturbative expansion of the Euclidean path integral) and the low temperature expansion (based on a ``long string effective theory''). The bootstrap results are seen to be consistent with, and interpolate between, the leading order analytic results in both limits. In the low temperature regime, the bootstrap bounds are sufficiently accurate to allow for extracting the energy gap $\Delta_1$ of the adjoint sector of the MQM.

The bootstrap method is also applicable to metastable thermal states at large $N$, as demonstrated in section \ref{sec:unbounded} for the Hamiltonian (\ref{1matH}) with a quartic potential that is unbounded from below. In this case, the bootstrap constraints are found to be unfeasible above a critical temperature $T_c$, where the thermal state ceases to exist.

Additionally, we have carried out a preliminary thermal bootstrap analysis of the ungauged two-matrix QM described by the Hamiltonian
\ie\label{2matH}
H = {\rm Tr} \Big[ {1\over 2} P_1^2 + {1\over 2} P_2^2 + V(X_1, X_2) \Big],
\fe
with a quartic interaction potential $V$. The setup of the SDP problem is essentially identical to that of the one-matrix QM, except that the basis of ${\cal B}$ consists of words made out of the letters $X_1, X_2, P_1, P_2$ is a much bigger set at a given word length. Rigorous albeit crude bootstrap bounds on $E(\B)$ in the planar limit, as well as future prospects, are discussed in section \ref{sec:discuss}.

\section{Thermal bootstrap}\label{sec:TB}

In this section, we review the formulation of the KMS optimization as an SDP problem introduced in \cite{Fawzi:2023fpg,Fawzi_2018}, and demonstrate it in the example of a quantum anharmonic oscillator.

In a quantum system with Hilbert space $\cH$, the thermal expectation value of an operator ${\cal O}$ at inverse temperature $\B$ is given by
\ie
\corr{ \cO } \equiv \tr_{\mathcal{H}}\left(\R_\B\cO \right) ,
\fe
where $\R_\B$ is the thermal density matrix, related to the Hamiltonian $H$ and the partition function $Z(\B)$ by 
\ie
\R_\B={1\over Z(\B)} e^{-\B H},~~~~~ Z(\B) = \tr_{\mathcal{H}}\left(e^{-\B H}\right).
\fe
An essential property of the thermal expectation values is the KMS condition \cite{1957JPSJ...12..570K,1959PhRv..115.1342M,1967CMaPh...5..215H}\footnote{The notion of thermal states and the KMS condition can be formulated more generally without assuming that $e^{-\B H}$ is of trace-class; see Definition 5.3.1 of \cite{Bratteli:1996xq}.}
\beq\label{kms}
\corr{\cO_1 \cO_2} = \corr{\cO_2\, e^{-\beta H}\, \cO_1\, e^{\beta H}}\,,
\eeq
for any pair of operators ${\cO}_1$ and ${\cO}_2$. The stationary state conditions (\ref{SDeq}) immediately follows. Even though (\ref{kms}) is a linear relation on the thermal expectation values, its non-locality with respect to Euclidean time presents an obstacle for implementing the bootstrap approach of \cite{Han:2020bkb, Lin:2023owt}. In \cite{Araki:1977px,Sewell1977,Fawzi:2023fpg}, the KMS condition is reformulated\footnote{The definition of thermal states in \cite{Araki:1977px,Sewell1977}, motivated by requiring certain local thermodynamical stability conditions, is a priori distinct from the KMS condition but in fact equivalent. The authors also noted that such a definition reduces to Dobrushin, Lanford, and Ruelle (DLR) equation \cite{Dobruschin1968TheDO,1969CMaPh..13..194L} in the classical case. DLR equation is linear in the thermal expectation values of local operators and has recently been used to bootstrap the classical Ising model on the lattice \cite{Cho:2022lcj,Cho:2023ulr}.} as the convex conditions (\ref{sqrPositivity}), (\ref{kmseqn}) together with (\ref{SDeq}), where all thermal expectation values involved are those of operators localized in Euclidean time.

\subsection{Semidefinite relaxation of matrix logarithm}
\label{sec:approxlog}

The reformulation of (\ref{kmseqn}) as an SDP problem \cite{Fawzi_2018, Fawzi_2021} begins with the approximation of the logarithm function
\ie\label{app:rmk}
\log(x) \approx r_{m,k}(x) := 2^k r_m(x^{1/2^k}),
\fe
where $(m,k)$ is a pair of positive integers, and $r_m(x)$ is a sum of degree 1 rational functions,
\ie
r_m(x)\equiv \sum_{j=1}^m w_j f_{t_j}(x),~~~~ f_t(x)\equiv {x-1\over t(x-1)+1}.
\fe
Here $t_j\in[0,1]$ and $w_j>0$ are the nodes and weights of the $m$-point Gauss-Radau quadrature, chosen so that $\int_0^1 p(t) dt = \sum_{j=1}^m w_j p(t_j)$ holds for any polynomial $p(t)$ of degree $2m-2$, with $t_1=0$.\footnote{There are different variants of the Gaussian quadrature that one may use to approximate $\log (x)$. A rigorous bootstrap requires using a quadrature that strictly bounds $\log (x)$ from above. Generic variants of the quadrature that approximates (and not necessarily bounds) $\log(x)$ may still be used for the thermal bootstrap, but the resulting bounds would no longer be rigorous and may be incompatible with the physical values.}

Two basic observations are as follows. First, with increasing $(m,k)$, $r_{m,k}(x)$ provide successively tighter {\it rigorous} upper bounds on $\log(x)$. Second, $f_t(x)$ is a concave function of $x$ at every $t\in[0,1]$, and admits the SDP representation
\ie\label{ftineq}
f_t(x)\geq\tau~~\Longleftrightarrow~~
{\begin{bmatrix}
x-1-\tau &~~ -\sqrt{t}\tau\\
-\sqrt{t}\tau &~~ 1-t\tau
\end{bmatrix}}~\succeq~0.
\fe
In generalizing this idea to the matrix logarithm, the concave function $r_{m,k}(x)$ will be promoted to a matrix function $r_{m,k}(X)$, where the matrix $X$ takes value in $H^n_+$, the space of positive-definite $n\times n$ Hermitian matrices.

The logarithm appearing in (\ref{kmseqn}) is a perspective transformation of the matrix logarithm function, also known as the negative operator relative entropy,
\ie
D(A,B):= - A^{1/2}\log\left(A^{-1/2}B A^{-1/2}\right)A^{1/2},~~~~A,B\in H^n_+.
\fe
The semidefinite inequality (\ref{kmseqn}) is equivalent to the statement that the matrices $(A,B, {\cal T}=\B C)$ lie in the relative entropy cone
\ie
K^n=\text{cl}\{(A,B,{\cal T})\in H^n_+\times H^n_+\times H^n:~{\cal T}\succeq D(A,B) \},
\fe
where 'cl' stands for the closure, and $H^n$ is the space of $n\times n$ Hermitian matrices. Generalizing (\ref{app:rmk}), one defines the relaxed relative entropy cone as
\ie\label{kmkcone}
K^n_{m,k}=\text{cl}\{(A,B,{\cal T})\in H^n_+\times H^n_+\times H^n:~{\cal T}\succeq D_{m,k}(A,B) \},
\fe
where
\ie
D_{m,k}(A,B) :=-A^{1/2}r_{m,k}\left(A^{-1/2} B A^{-1/2}\right) A^{1/2}.
\fe
The upper bound of logarithm by $r_{m,k}$ leads to the property $K^n_{m,k}\supset K^n$, i.e. the convex condition defined by $K^n_{m,k}$ is a rigorous relaxation of the condition defined by $K^n$. As $(m,k)$ increases, $K^n_{m,k}$ provides a closer approximation of $K^n$; the error of this approximation was investigated in detail in \cite{Fawzi_2018}. 

Generalizing (\ref{ftineq}), the following theorem (Theorem 3 of \cite{Fawzi_2018}) reformulates $K^n_{m,k}$, as an approximation and a rigorous relaxation of the thermal bootstrap condition (\ref{kmseqn}), in the form of an SDP problem:
\noindent $(A,B,{\cal T})\in K^n_{m,k}$ if and only if there exist a set of matrices $T_1,\cdots,T_m$, $Z_0,\cdots,Z_k\in H^n$ such that
\ie\label{app:SDPform}
&Z_0=B,~\sum_{j=1}^mw_jT_j=-2^{-k}{\cal T},~{\begin{bmatrix}
Z_i &~~ Z_{i+1}\\
Z_{i+1} &~~ A
\end{bmatrix}}~\succeq~0,~{\rm and}
~~{\begin{bmatrix}
Z_k-A-T_j &~~ -\sqrt{t_j}T_j\\
-\sqrt{t_j}T_j &~~ A-t_jT_j
\end{bmatrix}}~\succeq~0
\fe
for $i=0,\cdots,k-1$ and $j=1,\cdots,m$.

In other words, the variables of the SDP reformulation of (\ref{kmseqn}) are the original $A,B,{\cal T}=\B C$ that are subject to the linear constraints that follow from the stationary state conditions on thermal expectation values, together with the $m$ auxiliary matrices $T_j$ and $k$ auxiliary matrices $Z_i$ that are required to satisfy (\ref{app:SDPform}).

Before closing this subsection, it is worth mentioning that instead of the SDP relaxation adopted in this work, a convex optimization directly implementing (\ref{kmseqn}) using the interior-point method is also possible. The relevant self-concordant barrier function has been recently constructed in \cite{Fawzi:2022glt} and implemented on a solver \cite{He:2024dco}.\footnote{We thank Hamza Fawzi and Kerry He for relevant discussions.}

\subsection{Warm-up example: the anharmonic oscillator at finite temperature\label{warmup}}

As an illustration of the SDP algorithm as well as a test of the quadrature relaxation of matrix logarithm, we now illustrate the thermal bootstrap in the simple example of the quantum anharmonic oscillator, namely a system of 1 degree of freedom described by the Hamiltonian
\ie
\label{hamanh}
H = p^2 + x^4\,.
\fe
A basis of operators ${\cal O}_i$ is labeled by words consisting of the letters $x$ and $p$. We denote by ${\cal B}_\ell$ the truncated basis of all such operators up to maximal word length $\ell$. For instance, 
\ie
{\cal B}_2 = \{1, x, p, x^2, p^2, xp, px \}. 
\fe
The variables of the convex optimization problem will include the set of thermal expectation values
\ie\label{eldef}
{\cal E}_{L}:= \{\langle {\cal O}\rangle_\B|{\cal O}\in{\cal B}_{L} \},
\fe
which are subject to linear relations due to the canonical commutation relation $[x,p]=i$ and the stationary state conditions $\langle [H,{\cal O}]\rangle_\B = 0$. For instance, $\langle xp\rangle_\B-\langle px\rangle_\B=i$, and $0=\langle[H,x^2]\rangle_\B=-2-4i\langle xp\rangle_\B$.

Next, we introduce the matrix (for even $L$)
\ie
{\cal M}^{(L)}_{i'j'}\equiv \langle {\cal O}^\dagger_{i'} {\cal O}_{j'}\rangle_\B,~~~~{\cal O}_{i'},{\cal O}_{j'}\in {\cal B}_{L\over 2},
\fe
which is constrained to be positive-definite due to the positivity of the thermal density matrix. The KMS condition is reformulated in (\ref{kmseqn}) in terms of the matrices $A,B,C$, constructed as
\ie
A^{(L)}_{ij}=\langle {\cal O}^\dagger_i {\cal O}_j\rangle_\B,~~~~B^{(L)}_{ij}=\langle {\cal O}_j {\cal O}^\dagger_i \rangle_\B,~~~~C^{(L)}_{ij}=\langle {\cal O}^\dagger_i [H,{\cal O}_j]\rangle_\B,~~~~{\cal O}_{i}, {\cal O}_j\in{\cal B}_{{L\over2}-2}.
\fe
For the SDP reformulation of the KMS condition (\ref{app:SDPform}), additional variables are introduced through the auxiliary $|{\cal B}_{{L\over2}-2}|\times|{\cal B}_{{L\over2}-2}|$ Hermitian matrices $T_1,\cdots,T_m$ and $Z_0,\cdots,Z_k$ (i.e. $n=|{\cal B}_{{L\over2}-2}|$), where $|{\cal B}_{{L\over2}-2}|$ is the cardinality of the set ${\cal B}_{{L\over2}-2}$, and we take $A=A^{(L)}$, $B=B^{(L)}$, ${\cal T}=\B C^{(L)}$.

\begin{figure}[h!]
	\centering
	\includegraphics[width=0.5\linewidth]{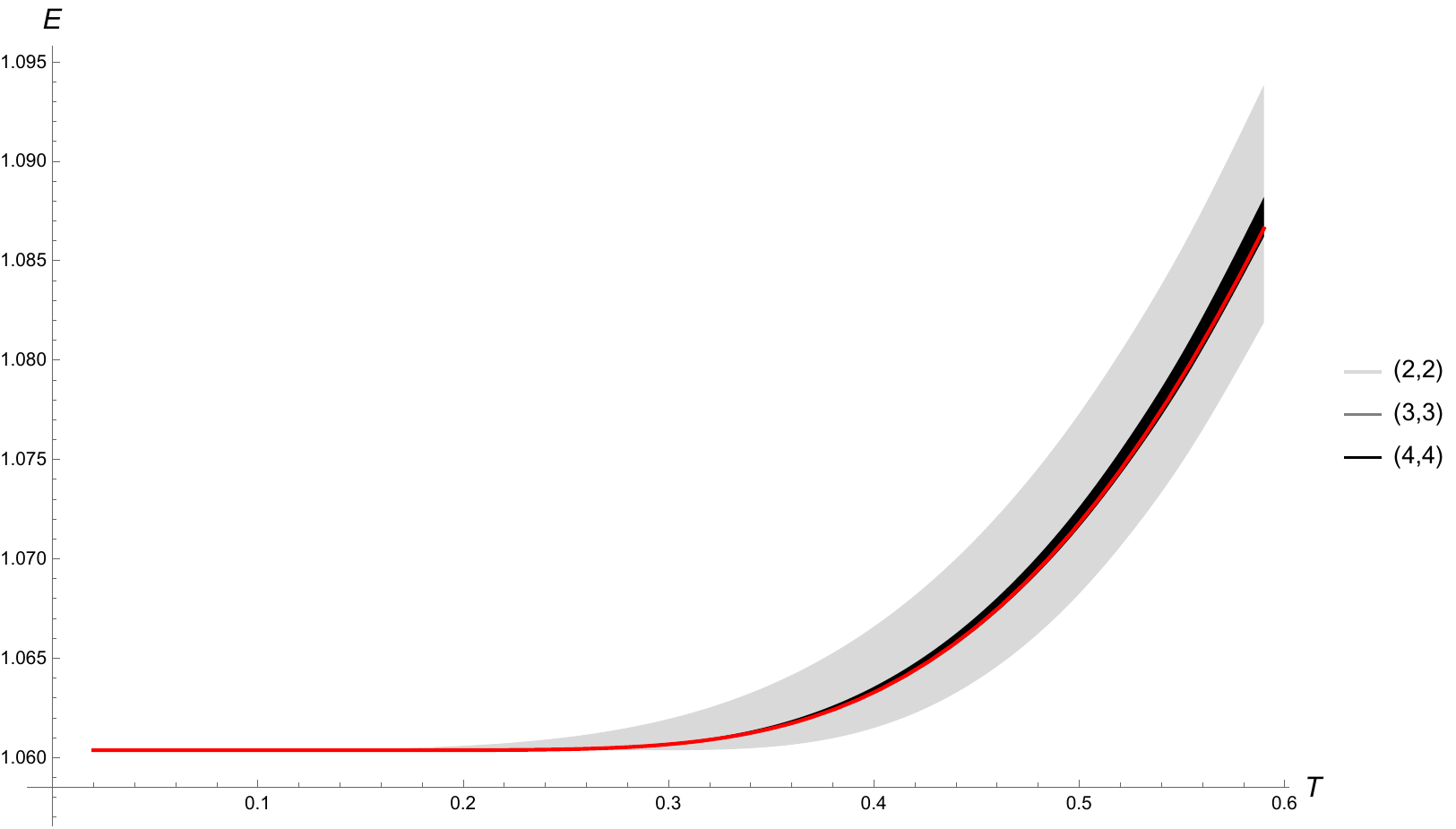}
	\caption{Bootstrap bounds on the thermal energy expectation value $E(\B)$ of the quantum anharmonic oscillator (\ref{hamanh}) as a function of the temperature $T=1/\B$, computed using different $(m,k)$-relaxations of the matrix logarithm, with maximal word length $L=10$.  Note that the bounds with $(m,k)=(3,3)$ and $(4,4)$ are virtually indistinguishable. The result from numerically solving the Schr\"odinger equation is shown in solid red curve. From the $(m,k)=(4,4)$ lower bootstrap bound, one can extract the energy gap $\Delta = 2.748$, which is within $\sim$ 0.3\% of the physical value $\Delta = 2.739$.}
	\label{fig:oneparticle}
\end{figure}

Given an objective Hermitian operator ${\cal O}$ which is a sum of words of length $\leq q$, a rigorous lower (upper) bound on the thermal expectation value $\langle{\cal O}\rangle_\B$ is obtained by solving the following SDP problem for any $L>{q}$.
\ie\label{SDP:single}
&\text{\textbf{SDP($L$)}: minimize (maximize) }\langle{\cal O}\rangle_\B\text{ subject to}
\\
&\text{1. Positivity of thermal density matrix: ${\cal M}^{(L)}\succeq~0$}
\\
&\text{2. Normalization: $\langle1\rangle_\B=1$}
\\
&\text{3. KMS condition: (\ref{app:SDPform}) with $n=|{\cal B}_{{L\over2}-2}|$, $A=A^{(L)}$, $B=B^{(L)}$, and ${\cal T}=\B C^{(L)}$}
\\
&\text{4. Canonical relation: $\langle {\cal O}_a [x,p] {\cal O}_b\rangle_\B=i\langle {\cal O}_a {\cal O}_b\rangle_\B$ for all ${\cal O}_a, {\cal O}_b$ such that ${\cal O}_a{\cal O}_b\in{\cal B}_{L-2}$}
\\
&\text{5. Stationary state conditions: $\langle[H,{\cal O}']\rangle_\B=0$ for all ${\cal O}'\in{\cal B}_{L-2}$}
\fe
The complete set of variables consists of ${\cal E}_L$ (\ref{eldef}) together with the auxiliary Hermitian matrices $T_1,\cdots,T_m$ and $Z_0,\cdots,Z_k$ that appear in the condition 3 of (\ref{SDP:single}). 

We have solved the SDP problem for the energy expectation value $E(\B) = \langle H\rangle_\B$ as the objective using SDPA-DD, for $L=10$ and different $(m,k)$-relaxations of the logarithm, with the results shown in Figure \ref{fig:oneparticle}. It is observed that at the given $L$, the upper and lower bounds already stabilize at $(m,k)=(3,3)$. Furthermore, the lower bound is close enough to the physical value to allow for extracting the energy gap between the ground state and the first excited state of the anharmonic oscillator to $\sim 0.3\%$ accuracy.

\section{The ungauged matrix quantum mechanics}\label{sec:themalmqm}

We are now turn to the main subject of this paper, the ungauged one-matrix QM defined by the Hamiltonian (\ref{1matH}), with the quartic potential
\ie\label{potentialspec}
V(X) = {1\over2}X^2+{g\over N}X^4.
\fe
The canonical coordinate $X$ and momentum $P$ are $N\times N$ traceless Hermitian matrices, whose components may be expressed as $P_{ab} = P^A T^A_{ab}$, $X_{ab} = X^A T^A_{ab}$, where $T^A_{ab}$ are the traceless Hermitian $SU(N)$ generators satisfying the completeness relation $T^A_{ab}T^{A}_{cd} = \delta_{ad}\delta_{bc} - \frac{1}{N}\delta_{ab}\delta_{cd}$. The canonical commutation relation $[P^A,X^B] = -i \delta^{AB}$ is equivalent to
\beq\label{commutator}
[P_{ab},X_{cd}] = -i \Big( \delta_{ad}\delta_{bc} - \frac{1}{N}\delta_{ab}\delta_{cd}\Big) .
\eeq
Note that this is slightly different from the Hermitian MQM where $X$ and $P$ are not subject to the traceless constraint, in which case the second (order ${1\over N}$) term on the RHS of (\ref{commutator}) is absent, although the two models share the same 't Hooft planar limit, i.e. $N\rightarrow\infty$ at fixed $g$. In this paper we focus on the traceless Hermitian MQM, which enjoys additional minor simplifications in the bootstrap setup.\footnote{The traceless constraint sets the single trace operators of length one, namely $\Tr X$ and $\Tr P$, to zero, and furthermore reduces the number of independent double trace expectation values after sorting words made out of $X$ and $P$ using commutation relations and cyclicity of the trace.}

The Hamiltonian (\ref{1matH}) admits a $U(N)$ global symmetry that acts by $X\rightarrow UXU^\dagger$, where $U$ is a unitary matrix, and therefore can be block-diagonalized according to irreducible representations of $U(N)$. The gauged version of the MQM, which amounts to restricting to the singlet sector of the Hilbert space, is well known to reduce to a system of $N$ identical non-relativistic fermions that are non-interacting apart from a constraint that is unimportant in the large $N$ limit,\footnote{The traceless property of $X$ amounts to constraining the sum of the $N$ fermion coordinates to zero.} as we briefly review in Appendix \ref{app:singlet}. In this work we will be interested in the ungauged MQM, where thermal observables receive contributions from the singlet as well as non-singlet representation sectors. The non-singlet sector dynamics does not reduce to that of non-interacting fermions and is nontrivial even in the large $N$ limit. The spectral problem in the adjoint representation, for instance, has been studied in \cite{Marchesini:1979yq}.

For the rest of this section, the objective of the thermal bootstrap is the energy expectation value at finite temperature $T=1/\B$,
\ie\label{energyexpect}
E(\B)\equiv \corr{H} = -{\partial \log Z(\B)\over\partial\B}.
\fe

\subsection{The SDP problem at finite $N$}
\label{sec:bstrap}

Extending the setup of section \ref{warmup} to MQM, we consider a basis of $U(N)$-adjoint-valued operators ${\cal O}_i$ constructed as the matrix product of an ``open string'' of $X$'s and $P$'s. We denote by ${\cal B}_\ell$ the truncated basis of all such operators that correspond to open string words up to maximal length $\ell$. For instance, 
\ie
\label{eq:openst2}
\mathcal{B}_2 =\{1,X,P,X^2,XP,PX,P^2\} ,
\fe
where each operator is understood as an $N\times N$ matrix.
The thermal two-point function of a pair of adjoint operators, say ${\cal O}_{ab}$ and ${\cal O}'_{cd}$, is decomposed according to $U(N)$-invariant tensor structures and reduces to the expectation values of fully traced operators,
\ie\label{SUNdecomp}
\corr{\cO_{ab}\cO'_{cd}} = \frac{1}{N^2-1}\Big( \delta_{ad}\delta_{bc} - \frac{1}{N}\delta_{ab}\delta_{cd}\Big)\Big[\corr{\Tr(\cO\cO')} &- \frac{1}{N}\corr{\Tr \cO\,\Tr \cO'}\Big] \\
&+\frac{1}{N^2} \delta_{ab}\delta_{cd} \corr{\Tr \cO\,\Tr \cO'} .
\fe
It is crucial here that we work with the ungauged theory, in which ${\cal O}_{ab}$ and ${\cal O}'_{cd}$ are valid linear operators acting on the Hilbert space.\footnote{In the gauged MQM, one would have to restrict ${\cal O}_i$ to traced operators, for which the thermal inequality (\ref{kmseqn}) trivializes in the planar limit due to the large $N$ factorization, e.g. $A_{ij} = B_{ij} + O(1/N)$.}
Let us also note that a priori, to optimize the bootstrap bounds one should also include ${\cal O}_i$ in multi-adjoint representations, but this appears unnecessary for the numerical implementation in this work.

As in the simple example of section \ref{warmup}, the expectation values of multi-trace operators are subject to \textit{linear} relations that follow from cyclicity of the trace, the reality condition $\langle\cO^\dagger\rangle_\B=\langle\cO\rangle^*_\B$,  the canonical commutation relation (\ref{commutator}), and the stationary state conditions $\langle [H, \Tr \cO\cdots \Tr \cO']\rangle_\beta=0$.\footnote{Note that in contrast to \cite{Han:2020bkb}, we do not have linear relations of the form $\langle{\rm tr} (G{\cal O})\rangle=0$ for $SU(N)$ generator $G$, due to the MQM being ungauged and that the thermal density matrix $\rho_\B$, although $SU(N)$-invariant, does not vanish when multiplied by $G$ from one side.} Simple examples of such linear relations are
\ie{}
& \corr{\Tr(PX)}-\corr{\Tr(XP)} = -i(N^2-1),
\\
& -i \corr{[H, \Tr(XP)]} = \corr{\Tr P^2}- \corr{\Tr X^2} -  \frac{4 g}{N} \corr{\Tr X^4} .
\fe
In our SDP implementation, the variables will include single-trace operators up to word length $L$, along with all multi-trace operators that appear in the $A,B,C$ matrices built out of adjoint ${\cal O}_i$'s (e.g. in (\ref{SUNdecomp})), as well as the multi-trace operators that arise through the stationary state conditions and the cyclicity relations that involve canonical commutators.

The positivity relations on thermal correlators can be organized according to the trace structures. In the finite $N$ MQM, we will work up to maximal word length $L=8$, in which case only two types of trace structures arise:
\ie
\mathcal{M}_{ij}^{(L,1)} = \corr{\Tr(\cO_i^\dagger \cO_j)}\ ,
\fe
corresponding to the exchange of adjoint representation between $\cO_i, \cO_j \in \mathcal{B}_{L/2}$, and
\begin{equation}
	\label{eq:posmatdt}
	\mathcal{M}_{ij}^{(L,2)} = \corr{\Tr(\cO_i^\dagger)\Tr( \cO_j)}\ ,
\end{equation}
corresponding to the exchange of singlet representation between $\cO_i, \cO_j$.

The matrices $A,B,C$ appearing in the KMS condition (\ref{kmseqn}) are a priori built out of the thermal expectation values of untraced operators. For instance, we can construct the matrix $A$ in terms of
\ie\label{aabcdij}
A^{(L) ab,cd}_{ij} = \corr{\cO_{i, ab}^\dagger \cO_{j, cd}}\,,
\fe
where $\cO_i,\cO_j\in\mathcal{B}_{{L\over2}-1}$, and similarly for the $B, C$ matrices. Of course, due to (\ref{SUNdecomp}), not all components of (\ref{aabcdij}) are independent. It suffices to work with the basis operators ${\cal O}_{i,11}$, ${\cal O}_{i,21}$, giving rise to 
\ie
\label{eq:1111}
A_{ij}^{(L) 11,11} &= \frac{1}{N(N+1)}\Big(\corr{\Tr(\cO_i^\dagger \cO_j)} + \corr{\Tr(\cO_i^\dagger)\Tr( \cO_j)}\Big),
\\
A_{ij}^{(L) 12,21} &= \frac{1}{N^2-1}\Big(\corr{\Tr(\cO_i^\dagger \cO_j)} -\frac{1}{N} \corr{\Tr(\cO_i^\dagger)\Tr( \cO_j)}\Big),
\fe
as well as the ``off-diagonal'' $A_{ij}^{(L) 11,21}$ and $A_{ij}^{(L) 12,11}$. In the practice, we find that including the off-diagonal components does not improve the thermal bootstrap bounds, and will therefore formulate the SDP problem by collecting the positivity constraints that involve the $\{ab,cd\}=\{11,11\}$, and separately the $\{12,21\}$, components of $A,B,C$.

\begin{figure}[h!]\
	\centering
	\includegraphics[width=0.9\linewidth]{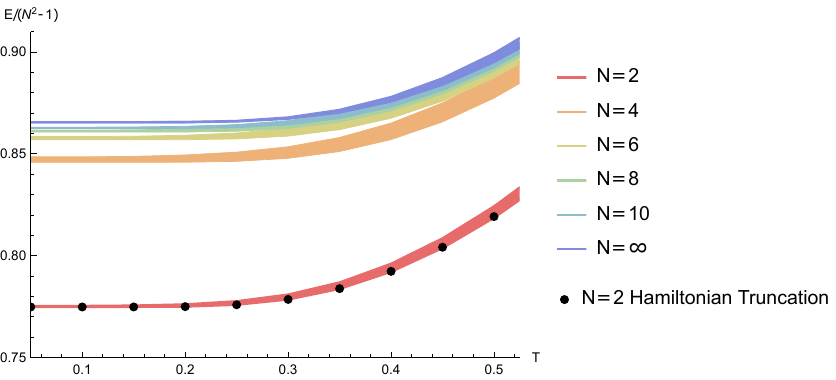}
	\caption{Bootstrap bounds on $E(\B)/(N^2-1)$ in the ungauged one-matrix QM at finite $N$ as well as at $N=\infty$, computed at coupling $g=2$ with maximal word length $L=8$, are shown in colors ranging from red to violet. For $N=2$, the Hamiltonian truncation results (shown in black dots) lie within the bootstrap bounds and differ from the lower bound by $\sim 10^{-4}$ at low temperatures to $\sim 10^{-2}$ at $T=0.5$.}
	\label{fig:nplot}
\end{figure}

Given an objective Hermitian operator ${\cal O}$ which is a sum of words of length $\leq q$, a rigorous lower (upper) bound on the thermal expectation value $\langle{\cal O}\rangle_\B$ is obtained by solving the following SDP problem for any $L>{q}$.
\ie\label{SDP:matrix}
&\text{\textbf{SDP($L$)} minimize (maximize) }\langle{\cal O}\rangle_\B\text{ subject to} 
\\
&\text{1. Positivity of thermal density matrix: ${\cal M}^{(L,i)}\succeq~0$}
\\
&\text{2. Normalization: $\langle1\rangle_\B=1$}
\\
&\text{3. KMS condition: (\ref{app:SDPform}) with $n=|{\cal B}_{{L\over2}-1}|$, $A=A^{(L)}$, $B=B^{(L)}$, ${\cal T}=\B C^{(L)}$}
\\
&\text{5. Cyclicity of the trace}
\\
&\text{6. Reality conditions}
\\
&\text{7. Canonical commutation relations}
\\
&\text{8. Stationary state conditions}
\fe
We have solved the SDP problem for the energy expectation value $E(\B)=\corr{H}$ as the objective using MOSEK, with the $(m,k)=(3,3)$ Gaussian quadrature approximation of the matrix logarithm, and working up to maximal word length $L=8$. The results for finite values of $N$ up to 10, as well as the infinite $N$ limit (see section \ref{sec:infnlim}), are shown in Figure \ref{fig:nplot}. The bounds on $E(\B)$ in the $N=2$ MQM is seen to be consistent with the results from Hamiltonian truncation.\footnote{The $N=2$ MQM is equivalent to a system of three coupled harmonic oscillators. We truncate the Hilbert space to the tensor product of three copies of the span of the first 14 energy eigenstates of a SHO, and then diagonalize the Hamiltonian of the coupled harmonic oscillator within this space.} Moreover, we observe that the lower bootstrap bound is in close agreement with the physical values determined from Hamiltonian truncation.

\subsection{The SDP problem in the infinite $N$ limit}\label{sec:infnlim}

It is of particular interest to perform the thermal bootstrap directly in the infinite $N$ or planar limit, where in addition to the constraints considered in (\ref{SDP:matrix}), the expectation values of multi-trace operators are further determined by those of the single trace operators via the large $N$ factorization,
\ie\label{largenfac}
\corr{\Tr(\cO_1)\cdots \Tr(\cO_n)} = \corr{\Tr(\cO_1)}\cdots\corr{\Tr(\cO_n)} \ .
\fe
While nonlinear constraints of the form (\ref{largenfac}) cannot be implemented directly in an SDP problem, they can be relaxed to linear semidefinite constraints as demonstrated in \cite{Kazakov:2021lel,Kazakov:2022xuh}. For our purpose, it is convenient to first reduce the number of independent multi-trace expectation values using the stationary state conditions,
\ie
\corr{\Tr(\cO_1)\cdots \Tr(\cO_{n-1})[H,\Tr(\cO_n)]} \to \corr{\Tr(\cO_1)}\corr{\Tr(\cO_2)}\cdots\corr{[H,\Tr(\cO_n)]} = 0 ,
\fe
and handle the remaining independent multi-trace variables using the relaxation method. In the examples we have encountered so far, the number of linearly independent multi-trace variables is either zero or small, and we will simply omit the nonlinear constraints in this work.

\begin{figure}[h!]
	\centering
	\includegraphics[width=1.
	\linewidth]{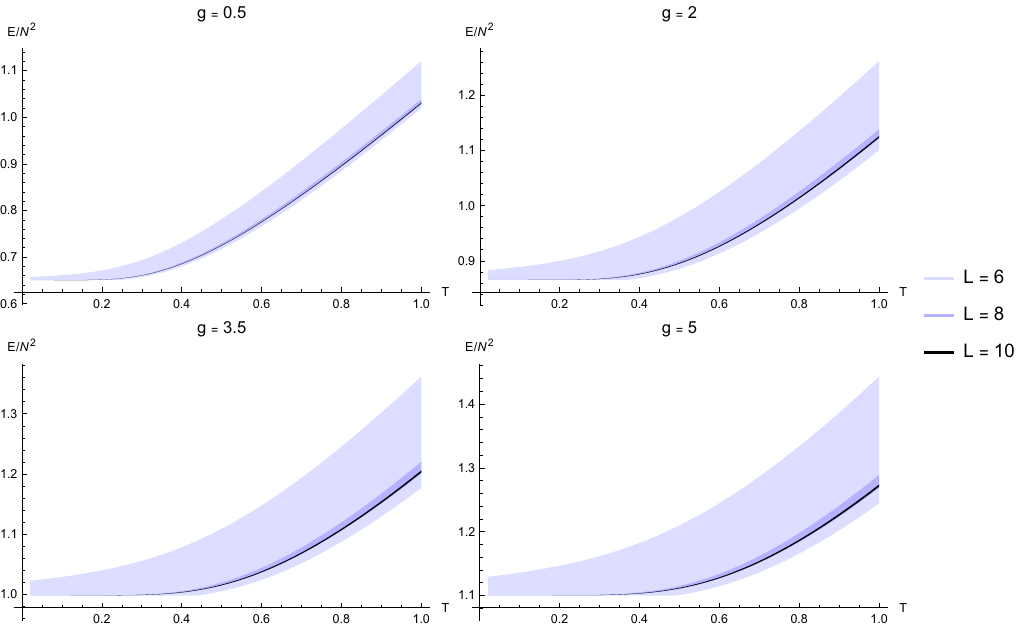}
	\caption{The bootstrap bounds on ${E(\B)/N^2}$ at various values of coupling $g$, computed using MOSEK with maximal word length $L=6$ and $L=10$, and using SDPA-DD with $L=8$ for numerical stability. The upper and lower bounds at $L=10$, shown in black, are virtually indistinguishable in the plot; their difference is further exhibited in Figure \ref{fig:convergence2}.}
	\label{fig:convergence}
\end{figure}

\begin{figure}[h!]
	\centering
	\includegraphics[width=1.
	\linewidth]{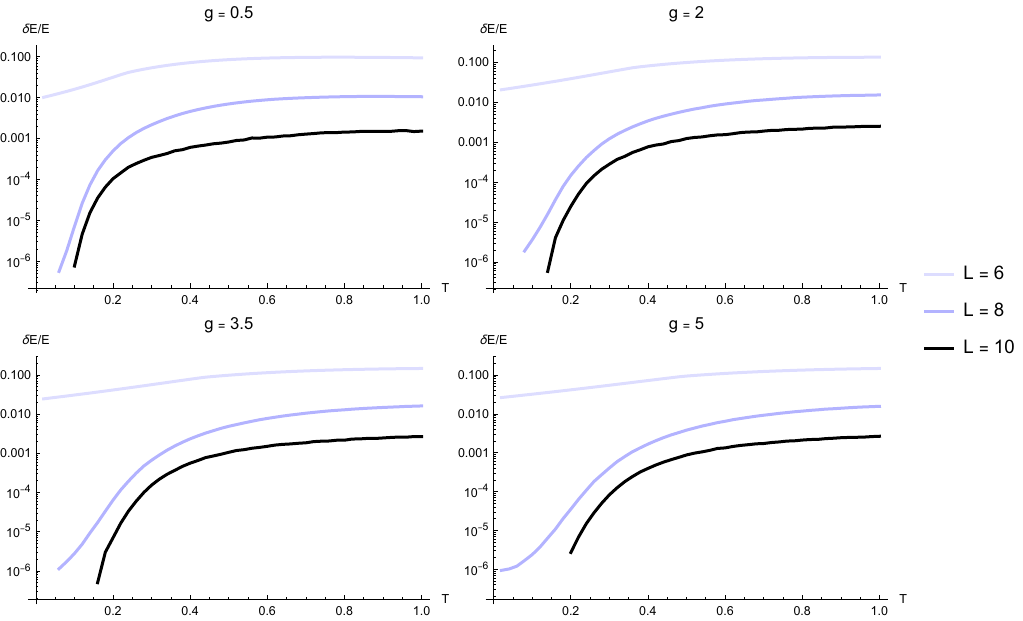}
	\caption{$\delta E/E$ as a function of the temperature, where $\delta E$ is the difference between the upper and lower bootstrap bounds on the energy expectation value $E$, is shown in logarithmic scale at various values of coupling $g$ and maximal word length $L$. 
	 }
	\label{fig:convergence2}
\end{figure}

The positivity relations are subject to a number of simplifications in the planar limit. First, the matrix ${\cal M}^{(L,2)}$\eqref{eq:posmatdt} becomes trivially positive and its correspinding positivity condition can be omitted. Next, the matrices $A^{(L)11,11}$ and $A^{(L)12,21}$ (\ref{eq:1111}) reduce to
\ie
A_{ij}^{(L) 11,11} &\to  \frac{1}{N^2}\Big( \corr{\Tr(\cO_i^\dagger)}\corr{\Tr( \cO_j)}\Big) ,
\\
A_{ij}^{(L) 12,21} &\to \frac{1}{N^2}\Big(\corr{\Tr(\cO_i^\dagger \cO_j)} -\frac{1}{N} \corr{\Tr(\cO_i^\dagger)}\corr{\Tr( \cO_j)}\Big).
\fe
Note that each trace comes with a factor of $N$. Similar simplifications occur for the $B,C$ matrices. In fact, we observe that the matrix $C^{(L) 11,11}$ vanishes in the planar limit, as
\ie
C_{ij}^{(L) 11,11} \to  \frac{1}{N^2}\Big( \corr{\Tr(\cO_i^\dagger)}\corr{\Tr([H, \cO_j])}\Big) = 0.
\fe
In contrast, $C^{(L) 12,21}$ remains nontrivial in the planar limit,
\ie
C_{ij}^{(L) 12,21} &\to  \frac{1}{N^2}\Big(\corr{\Tr(\cO_i^\dagger [H, \cO_j])} -\frac{1}{N} \corr{\Tr(\cO_i^\dagger)}\corr{\Tr( [H, \cO_j])}\Big) 
\\
&=  \frac{1}{N^2}\corr{\Tr(\cO_i^\dagger [H, \cO_j])} .
\fe
Consequently, the $\{11,11\}$ component of the KMS condition (\ref{kmseqn}) has no nontrivial temperature dependence in the planar limit and can be omitted, leaving the $\{12,21\}$ component as the only nontrivial KMS condition that enter the SDP problem (\ref{SDP:matrix}).

Using these simplified positivity and KMS conditions in the planar limit, we have implemented the SDP problem (\ref{SDP:matrix}) with $E(\B)=\corr{H}$ as the objective, using the $(m,k)=(3,3)$ Gauss-Radau quadrature relaxation of matrix logarithm and working up to maximal word length $L=10$. The resulting upper and lower bootstrap bounds are shown in Figure \ref{fig:convergence} at different values of the 't Hooft coupling $g$. The gap between the upper and lower bounds tightens dramatically as $L$ increases, as exhibited in more detail in Figure \ref{fig:convergence2}.

\section{Comparison to analytic results}\label{sec:UMQM}

We now describe the analytic approach to the thermodynamics of the ungauged one-matrix quantum mechanics (\ref{1matH}), based on the conventional perturbative expansion of the functional integral in the high temperature limit, and based on the long string effective theory \cite{Gross:1990md,Maldacena:2005hi,Balthazar:2018qdv} in the low temperature limit.

The thermal partition function $Z(\B)$ of the matrix QM can be formulated in terms of the Euclidean path integral over the thermal circle \cite{Zinn-Justin:2000ecv, Kawahara:2007fn, Kawahara:2007ib},
\ie\label{pathInt}
Z(\B)=\int DX\exp\left( -\int_0^\B d\tau \, \Tr\left[{1\over2}(\partial_\tau X)^2+{1\over2}X^2+{g\over N}X^4 \right]\right),
\fe
where $X(\tau)$ is subject to the periodic boundary condition $X(\tau+\B)=X(\tau)$. The energy expectation value (\ref{energyexpect}) can be expressed using Virial relation as
\ie\label{EpathInt}
E(\B)={1\over\B}\bigg\langle \int_0^\B d\tau\left({3g\over N}\Tr X(\tau)^4+\Tr X(\tau)^2\right)\bigg\rangle.
\fe
In the high temperature (small $\B$) regime, (\ref{EpathInt}) can be evaluated in perturbation theory. Expanding $X(\tau)$ in its Fourier modes $X(\tau)={1\over\sqrt{\B}}\sum_{m}X_m e^{2\pi im\tau/\B}$, the kinetic term of the action reads $\sum_m{1\over2}({2\pi m\over\beta})^2\Tr\left(X_{-m}X_m\right)$. In the $\B\rightarrow0$ limit, only the zero mode $X_0$ is relevant, and the functional integral of (\ref{EpathInt}) reduces to the matrix integral in $X_0$,
\ie\label{highTE}
E(\B)={\int dX_0 \, e^{-\Tr\left( {1\over2}X_0^2+{g\over \B N}X_0^4 \right) } \left({3g\over\B^2 N}\Tr X_0^4+{1\over\B}\Tr X_0^2\right) \over\int dX_0 \,e^{-\Tr\left( {1\over2}X_0^2+{g\over \B N}X_0^4 \right) }  }+{\cal O}(\B^{1\over 2}).
\fe
The exact solution of this matrix model is well known \cite{DiFrancesco:1993cyw}. In the planar limit, for $g>0$ (\ref{highTE}) evaluates to\footnote{Note that while a naive power counting might have suggested that the MQM is weakly coupled at high temperatures, the model at nonzero coupling $g$ has a different $\B\to 0$ asymptotic behavior from the Gaussian model at $g=0$. } (see e.g. (3.7) of \cite{Kazakov:2021lel})
\ie\label{gN0highT}
E(\B) = {N^2\over \B} \left[ {3\over 4} - \C - {2\over 3}\C^2 + {2\over 3} \C^{1\over 2} (1+\C)^{3\over 2} \right] +{\cal O}(\B^{1\over 2}),~~~~ \C \equiv {\B \over 48 g}.
\fe
The leading correction to (\ref{gN0highT}) comes from a planar diagram that is 1-loop with respect to the nonzero modes $X_{m\not=0}$, which evaluates to ${2\over 3\sqrt{3}} N^2 (\B g)^{1\over 2}$.

\begin{figure}[h!]
	\centering
	\includegraphics[width=1.
	\linewidth]{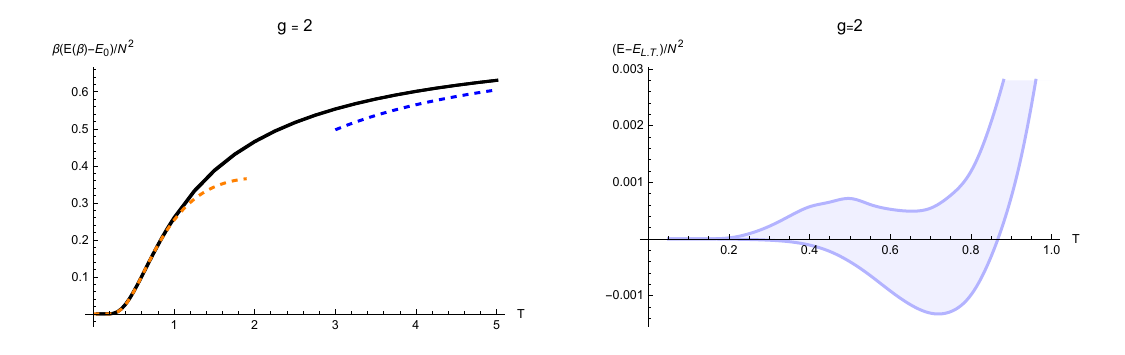}
	\centering
	\includegraphics[width=1.
	\linewidth]{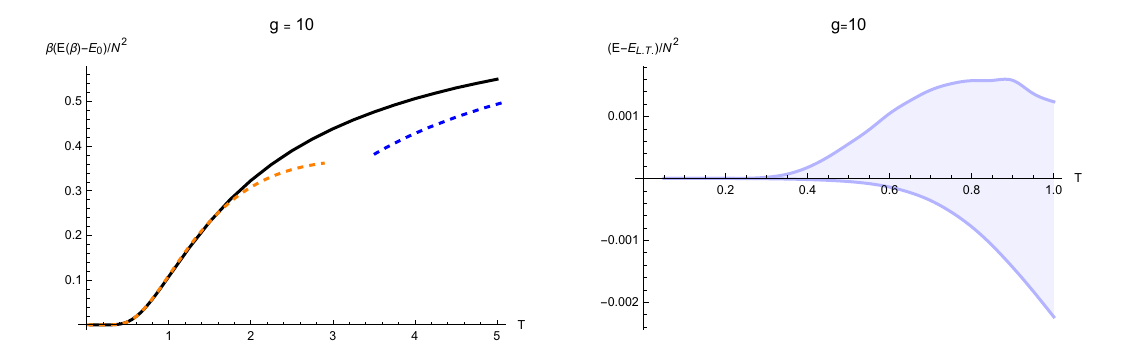}
	\centering
	\includegraphics[width=1.
	\linewidth]{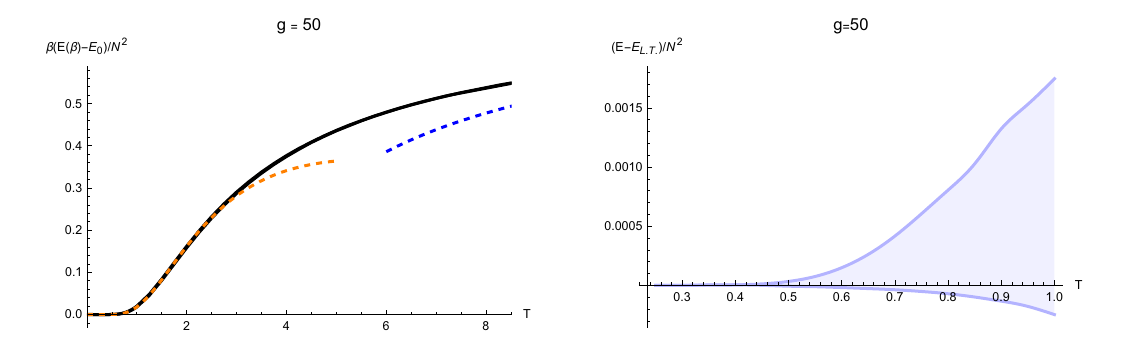}
	%	\centering
	\caption{Left: bootstrap results for $\B(E(\B)-E_0)/N^2$ in the ungauged one-matrix QM computed with maximal word length $L=10$, shown in solid black curve (where the distinction between the upper and lower bounds is invisible), compared to the leading terms in the low temperature expansion (\ref{lowTExp}) shown in dashed orange, and the leading terms in the high temperature expansion (\ref{gN0highT}) shown in dashed blue.
Right: the upper and lower bootstrap bounds on $E(\B)/N^2$ after subtracting off the low temperature approximation $E_{L.T.}(\B)/N^2 = e_0 + \Delta_1 e^{-\B\Delta_1}$. 
	}
	\label{fig:lowhighexp}
\end{figure}

To understand the low temperature regime, we begin with the decomposition of the thermal partition function according to representations of $SU(N)$,
\ie\label{thermalPF}
Z(\B)=\sum_{R}\sum_{s_R}\exp(-\B E_{s_R}),
\fe
where the first sum is taken over all representations $R$ obtained from the symmetric products of the adjoint (including the singlet), and the second sum is taken over energy eigenstates $|s_R\rangle$ in the representation $R$, of energy $E_{s_R}$.
At zero temperature ($\B=\infty$), the sum on the RHS (\ref{thermalPF}) reduces to the contribution from the ground state which belongs to the $SU(N)$ singlet sector. The latter admits a well-known free fermion description (see Appendix \ref{app:singlet}), in which the ground state corresponds to the fermion sea filling the $N$ lowest energy levels. In the large $N$ limit, the ground state energy is
\ie\label{gsE}
E(\beta=\infty)=E_0 = e_0 N^2,
\fe
where $e_0$ is an order 1 number (see e.g. (\ref{SingAdjE})).

The key observation \cite{Gross:1990md,Maldacena:2005hi,Maldacena:2018vsr,Balthazar:2018qdv} is that the low lying states of the adjoint sector admit the effective description in terms of a ``long string'',\footnote{The terminology refers to the holographic interpretation of the adjoint sector of a large $N$ gauge theory as a long open string whose ends are anchored to classical quark sources on the asymptotic boundary. In particular, the quartic interaction appearing in the effective Hamiltonian (\ref{lsham}) can be understood as a long strings reconnecting in the bulk.} and that the multi-adjoint sectors are described by multiple long strings that are weakly interacting with one another in the low temperature limit. As reviewed in Appendix \ref{app:singleadj}, the energy eigenstates $|n_{\rm adj}\rangle$ in the adjoint sector are characterized by the long string wave function $w_n(\lambda)$, where $\lambda$ is the coordinate on the configuration space of matrix eigenvalues. The corresponding long string creation and annihilation operators are denoted $(\mathfrak{a}_n^\dagger)_{ab}$ and $(\mathfrak{a}_n)_{ba}$. The leading interaction between a pair of long strings is characterized by the effective Hamiltonian
\ie\label{lsham}
{\cal H}_{ls} = \sum_n \Delta_n \Tr (\mathfrak{a}_n^\dagger \mathfrak{a}_n) + {1\over 2N} \sum_{n,m,k,\ell} h_{k\ell nm}\sum_{a,b,c,d}\left(\mathfrak{a}_n^\dagger\right)_{ab} \left(\mathfrak{a}^\dagger_m\right)_{cd} \left(\mathfrak{a}_k\right)_{bc} \left(\mathfrak{a}_\ell\right)_{da},
\fe
where the coupling coefficients $h_{k\ell nm}$ (\ref{hcoeff}) is determined in Appendix \ref{app:adjint} from the bi-adjoint sector of the MQM.

The effect of the long string interaction is most easily understood by reformulating (\ref{lsham}) in terms of the path integral based on the Euclidean action
\ie\label{seeuc}
S_E = \int_{-{\B\over 2}}^{\B\over 2} d\tau\, \Tr\left[ \sum_n \phi_n^* \left(\partial_\tau + \Delta_n\right)\phi_n + {1\over 2N} \sum_{n,m,k,\ell} h_{k\ell nm}\phi_n^{*}\phi_k\phi_m^{*}\phi_\ell \right] ,
\fe
where $(\phi_n)_{ab}(\tau)$ are adjoint-valued field variables on the thermal circle parameterized by $\tau \in [-\B/2, \B/2]$. The propagator for $\phi_n$ is given by the Green function
\ie\label{tauprop}
G_{nm}(\tau)=\delta_{nm} {\exp\left(-\Delta_n\tau-{\B\Delta_n\over2}\right)\over1-\exp(-\B\Delta_n)}\exp\left({\B\Delta_n\over2}\text{sgn}(\tau)\right),
\fe
where $\text{sgn}$ is the sign function, and the adjoint indices are suppressed. Note that (\ref{tauprop}), by construction, satisfies the periodic boundary condition at $\tau=\pm\B/2$. On the other hand, there is a discontinuity of $G(\tau)$ at $\tau=0$, and in particular $G(0^-)$ and $G(0^+)$ are distinct. A self-contraction on a vertex in a Feynman diagram is computed using $G_{nm}(0^-)={\delta_{nm}\over \exp(\B\Delta_n) - 1}$, as dictated by the ordering prescription of (\ref{lsham}). The long string thermal partition function in the planar limit up to 2-loop order is computed as
\ie
\log Z(\B) = N^2\left[ -\sum_n \log (1-e^{-\B \Delta_n}) - \B \sum_{n,m} {h_{nmnm} \over (e^{\B\Delta_n}-1)(e^{\B\Delta_m}-1)} 
+{\cal O}(h^2) \right].
\fe
Note that the corrections due to long string interactions are suppressed in the low temperature limit, as $G(\tau)$ is supported at positive $\tau$ only and all planar graphs constructed from the interaction vertex of (\ref{seeuc}) are suppressed. We thus find the low temperature expansion of the energy expectation value in the planar limit, including the ground state energy (\ref{gsE}),
\ie\label{lowTExp}
{E(\B)\over N^2}=e_0+\Delta_1\exp(-\B\Delta_1)+ {\cal O}(e^{-\B\Delta_2}, e^{-2\B\Delta_1}),
\fe
where $\Delta_1$ is the lowest energy eigenvalue of a long string (\ref{SingAdjE}), and the omitted terms include contributions from the excited long strings as well as multi-long string states.

In Figure \ref{fig:lowhighexp} the bootstrap bounds on $E(\B)$ are seen to interpolate between the leading order results in the high temperature expansion (\ref{gN0highT}) and in the low temperature expansion (\ref{lowTExp}), and is consistent with both in the respective high and low temperature limits. In the low temperature regime, the bootstrap bounds are sufficiently tight to allow for the extraction of the energy gap $\Delta_1$ of the adjoint sector that appears in (\ref{lowTExp}), giving results within $0.5\%$ ($3\%$) of the physical value (\ref{SingAdjE}) when fit to the lower (upper) bound. 
In the intermediate temperature regime, the bootstrap has provided numerical solutions beyond what is easily accessible by the analytic methods.

\section{Thermal phase transition at large $N$}\label{sec:unbounded}

\begin{figure}[h!]
	\centering
	\includegraphics[width=.9
	\linewidth]{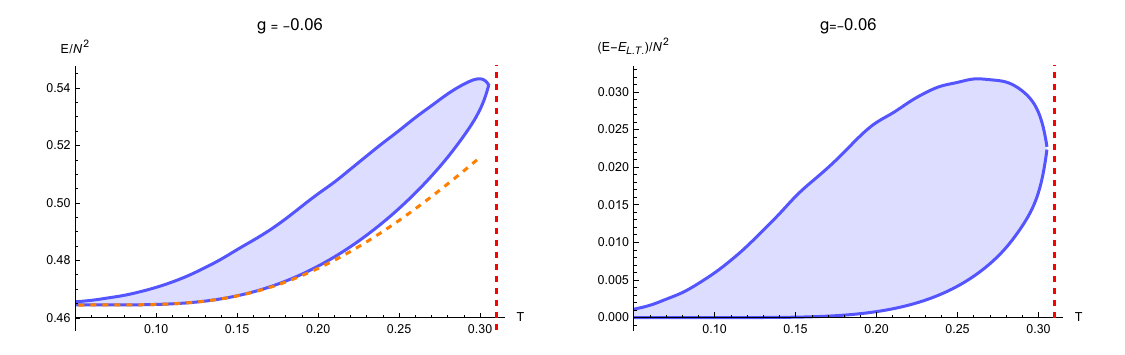}
	\includegraphics[width=.9
	\linewidth]{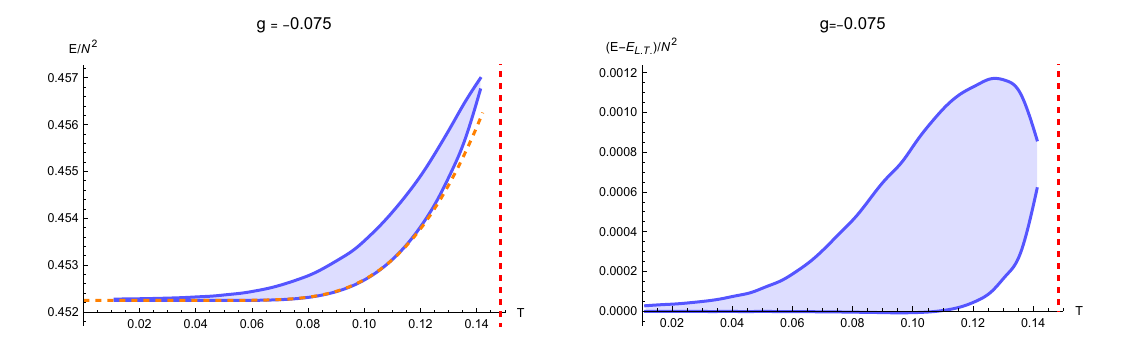}
	\caption{Left: the domain of $E(\B)/N^2$ (shaded light blue) allowed by the upper and lower bootstrap bounds (solid blue) at various values of negative coupling $g$, computed with maximal word length $L=10$. The low temperature expansion result, i.e. $E_{L.T.}(\B)/N^2$, is shown in dashed orange. Right: the allowed domain of $(E(\B)-E_{L.T.}(\B))/N^2$. The dashed red vertical line marks $T=T^{SDP}_c$ beyond which the bootstrap SDP is found to be infeasible (using SDPA-DD), indicating that no KMS state exists at temperatures above $T^{SDP}_c$.
}
	\label{fig:negativeg}
\end{figure}

\begin{figure}[h!]
	\centering
	\includegraphics[width=.5
	\linewidth]{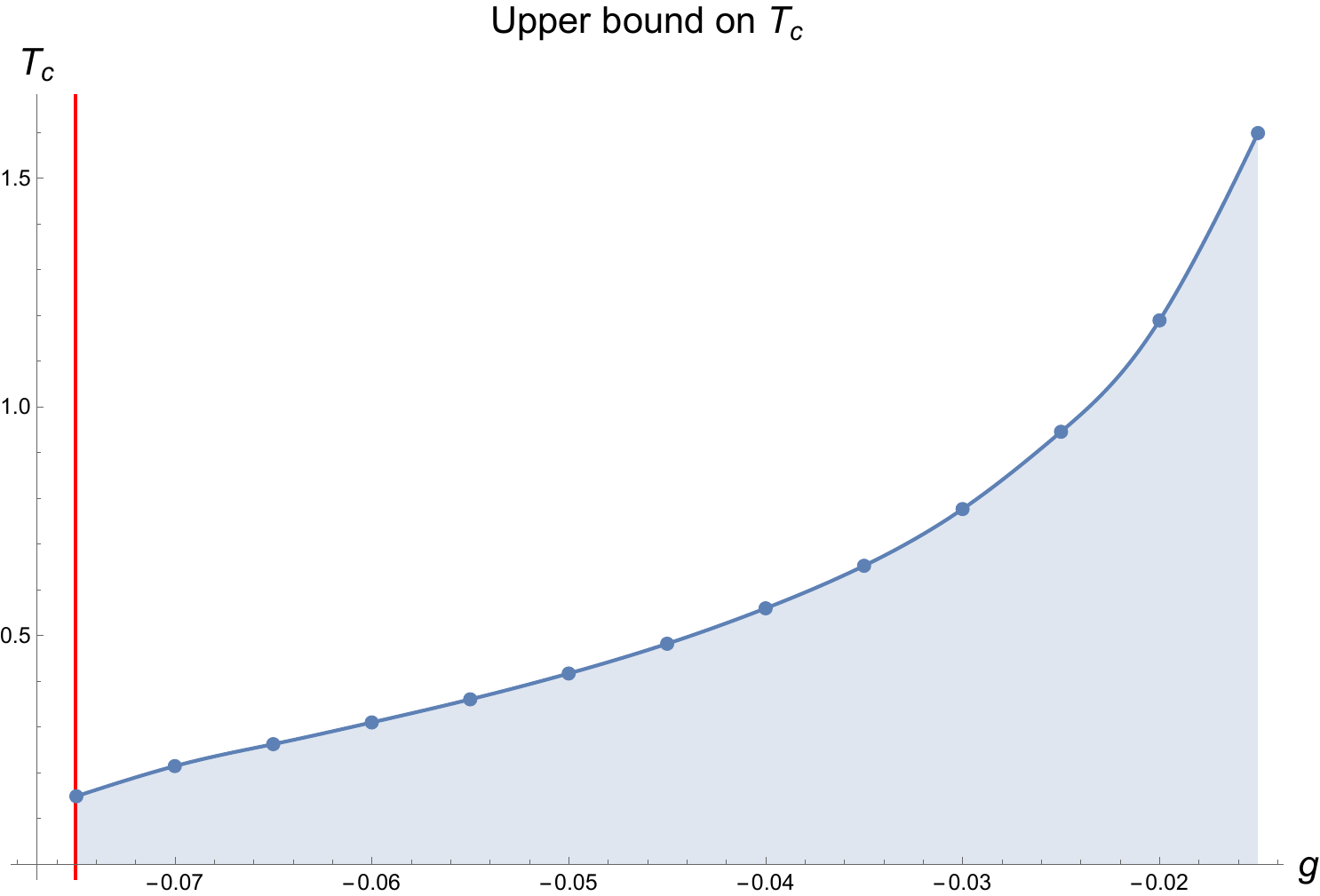}
	\caption{Upper bounds on the critical temperature $T_c$ as a function of the (negative) coupling $g$, computed using SDPA-DD at maximal word length $L=10$. The red vertical line marks $g=g_c$ at which $T_c$ vanishes.}
	\label{fig:negativegupper}
\end{figure}

Even for a Hamiltonian that is unbounded from below, there can be metastable states that become stable in the planar limit. For instance, the system (\ref{1matH}) with potential (\ref{potentialspec}) admits a singlet ground state so long as the quartic coupling $g$ lies in the range $g\geq g_c\equiv -{\sqrt{2}\over6\pi}$. We can ask if the analogous thermal state at a given temperature $T$ exists in the planar limit, for $g$ in the range $g_c\leq g<0$.

At $T=0$, the thermal state reduces to the ground state and thus exists for $g\geq g_c$. Exactly at $g=g_c$, the singlet ground state is described by a Fermi sea, whose Fermi energy is equal to the maximum of the potential $V$. In this case, a nonzero temperature would lead to the fermions spilling over the potential barrier, leading to an instability. We therefore expect that the thermal state with $T>0$ does not exist at $g=g_c$. For $g_c<g<0$, a thermal state may exist for nonzero temperature $T$ below a certain critical value $T_c(g)$. $T_c(g)$ must be finite, as the high temperature limit is characterized by the matrix integral (\ref{highTE}) which is unstable for any negative $g$.

Figure \ref{fig:negativeg} shows the bootstrap bounds on $E(\B)$ for $g<0$. In the SDP formulation of the thermal bootstrap, we can ask if the bootstrap constraints at a given $g<0$ are feasible or infeasible at some temperature $T=T^{SDP}$. If the constraints are found to be infeasible, we conclude that $T_c\leq T^{SDP}$. As such, the SDP provides a rigorous upper bound on the critical temperature $T_c$. This infeasibility test is performed at various values of $g<0$ and $T$ using SDPA-DD. 

Figure \ref{fig:negativegupper} shows the upper bound on the critical temperature $T_c^{SDP}$, that is the minimal value of $T=T^{SDP}$ at which the bootstrap constraints are found by SDPA-DD to be infeasible (by producing \textbf{phase.value = pdINF}), scanned over negative values of $g$. An interesting observation is that bootstrap bounds in Figure \ref{fig:negativeg} loosens as the temperature increases from zero, but tightens again as the temperature approaches $T^{SDP}_c$.

\section{Future prospects}
\label{sec:discuss}

We have thus far applied the thermal bootstrap, based on stationary state conditions and the KMS condition reformulated as an SDP problem through quadrature relaxations of the matrix logarithm, to the ungauged one-matrix QM and have obtained reasonably tight bounds on the energy expectation value at finite temperature that go beyond what is easily accessible through analytic methods in both the high temperature expansion and the low temperature expansion. The bootstrap bounds are tight enough in the low temperature regime to match with the leading correction in the low temperature expansion (\ref{lowTExp}) due to the lowest energy long string state in the adjoint sector of the MQM, but not yet tight enough to distill the excited long string states and the effects due to interactions between long strings. The latter is an interesting target for improving the efficiency and precision of the thermal bootstrap algorithm in the near future.

\begin{figure}[h!]
	\centering
	\includegraphics[width=.7
	\linewidth]{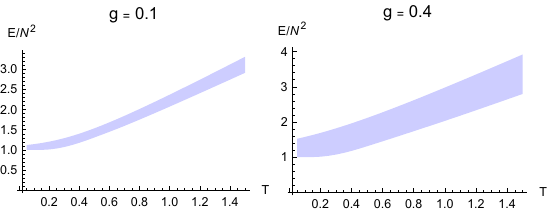}
	\caption{The allowed region of the thermal expectation value $E(\B)/N^2$ in the two-matrix QM with potential (\ref{eq:VYM}), computed from the bootstrap bounds at maximal  word length $L=4$. }
	\label{fig:YM4}
\end{figure}

The thermal bootstrap is of course not limited to the one-matrix QM. This is readily demonstrated by a preliminary bootstrap analysis of the two-MQM system (\ref{2matH}), consisting of Hermitian matrix valued coordinates $X_1, X_2$ and their conjugate momenta $P_1, P_2$, interacting through the potential 
\ie
\label{eq:VYM}
	V(X_1,X_2) = \frac{1}{2}\left(X_1^2 + X_2^2\right)-\frac{g}{N} [X_1,X_2]^2\ .
\fe
The setup of the SDP problem is identical to that of section \ref{sec:themalmqm}, starting with the truncated basis of adjoint operators $\mathcal{B}_\ell$ that contains open string words made out of the letters $X_1, X_2, P_1, P_2$ up to length $\ell$.
For instance, \eqref{eq:openst2} is now replaced with
\ie
\mathcal{B}_2 &=\{1,X_1,X_2,P_1,P_2,X_1^2,X_1 X_2,X_2 X_1,X_2^2,
X_1 P_1,P_1X_1,
\\
&~~~~~ X_1 P_2, P_2 X_1,X_2 P_1,P_1 X_2, 
X_2 P_2, P_2 X_2,P_1^2, P_1 P_2, P_2 P_1,P_2^2 \}.
\fe
We have implemented the SDP problem (\ref{SDP:matrix}) with the objective $E(\B)=\corr{H}$ in the planar limit, working up to maximal word length $L=4$ for the thermal expectation values appearing in the $A,B,C$ matrices. The resulting bounds as functions of the temperature are shown in Figure \ref{fig:YM4}. 

Evidently, the set of basis operators up to a given word length $L$ grows rapidly with the number of matrix canonical variables in a multi-matrix QM, posing a daunting challenge for the thermal bootstrap SDP setup. On the other hand, preliminary analysis of the 2-matrix QM suggests that meaningful bootstrap bounds in a multi-matrix QM may be obtained without necessarily going to high values of $L$, so long as the truncated basis of operators is itself reasonably large. It remains to be seen whether the thermodynamics of holographic MQM such as the (ungauged \cite{Maldacena:2018vsr}) BFSS model is tractable in the bootstrap approach. 

An unresolved challenge for the bootstrap method developed in this work is the thermodynamics of gauged MQM. In contrast to the ungauged MQM in section \ref{sec:themalmqm} where it was crucial that the ${\cal O}_i$'s are untraced operators corresponding to open string words, in a gauged system the matrices $A,B,C$ (\ref{sqrPositivity}), (\ref{ABCmats}) can only be built out of gauge-invariant basis operators ${\cal O}_i$, and the KMS inequality (\ref{kmseqn}) trivializes in the planar limit due to large $N$ factorization. One possible workaround is to implement the gauging not by restricting the Hilbert space to the singlet sector directly, but by deforming the ungauged MQM Hamiltonian so as to lift the energy of the non-singlet sectors. However, this introduces additional complexity into the Hamiltonian and the stationary state conditions, and its numerical effectiveness remains unclear at the moment. On the other hand, in holographic theories the non-singlet sectors may be lifted by the strong coupling dynamics at large $N$ \cite{Maldacena:2018vsr}, in which case gauging (for the bootstrap) may be unnecessary after all.

\section*{Acknowledgements}

We would like thank Hamza Fawzi, Sean Hartnoll, Kerry He, Antal Jevicki, Henry Lin, Martin Kruczenski, Subir Sachdev, Toby Wiseman and Zechuan Zheng for discussions. We thank the programs ``What is String Theory? Weaving Perspectives Together'' at Kavli Institute for Theoretical Physics, Santa Barbara, and Bootstrap 2024 at Complutense University, Madrid during the course of the work. This work is supported in part by DOE grant DE-SC0007870, by the Simons Collaboration on Confinement and QCD Strings, and by grant NSF PHY-2309135 to the Kavli Institute for Theoretical Physics (KITP). The work of MC was supported in part by the Sam B. Treiman fellowship at the Princeton Center for Theoretical Science.  The numerical computations in this work are performed on the FAS Research Computing cluster at Harvard University.

\appendix

\section{SDP solvers}

In this work, we use two SDP solvers to implement the thermal bootstrap at the numerical level. The first is the double precision solver MOSEK \cite{mosek} with default parameter values. The second is the double-double precision solver SDPA-DD \cite{5612693,sdpa,sdpaManual},\footnote{We used the version of SDPA-DD available at \cite{sdpaddweb}.} with following parameter values: epsilonStar=1.0E-0, lambdaStar=1.0E3, omegaStar=2.0, lowerBound=-1.0E5, upperBound = 1.0E5, betaStar=0.1, betaBar=0.2, gammaStar=0.9, epsilonDash=1.0E-9.\footnote{The vanilla SPDA solver was used for finite $N$ results with parameter values: epsilonStar=1.0E-6, betaStar=0.1, betaBar=0.2, gammaStar=0.5, epsilonDash=1.0E-6.}   Unless otherwise mentioned, we use MOSEK to produce upper and lower bounds on the finite temperature observables since it is much faster than SDPA-DD. However, there are cases where MOSEK is numerically unstable and does not find optimal solutions. In such cases, we use the higher precision solver SDPA-DD to find either bounds or infeasibility certificates. We explicitly mention whenever SDPA-DD is used. MOSEK was run on a single-core FAS Research Computing cluster at Harvard University, and SDPA-DD was run on a 10-core Intel i9-10900F processor. Sample computing times for these solvers are displayed in the following tables.

\begin{table}[H]
	\centering
	\caption{Anharmonic oscillator at $L=10$ in SDPA-DD (Section \ref{warmup}). The SDP problem had 10 variables after the commutation relations and stationary state conditions were imposed, in addition to the auxiliary matrices $T_j$ and $Z_i$. The positivity matrix was $63 \times 63$ and the $A$, $B$, $C$ and auxiliary matrices were $16 \times 16$.}
	\begin{tabular}{|c|c|}
		\hline
		\textbf{Setup} & \textbf{Runtime} \\ \hline
		$(m,k) = (2,2)$ & $\sim$ a few minutes  \\ \hline
		$(m,k) = (3,3)$ & $\sim$ 10 minutes  \\ \hline
		$(m,k) = (4,4)$ & $\sim$ 20 minutes  \\ \hline
	\end{tabular}
\end{table}

\begin{table}[H]
	\centering
	\caption{One-matrix quantum mechanics at $N=\infty$ (Section \ref{sec:infnlim}). The $L=10$ SDP problem had 175 variables after taking the cyclicity of trace, commutation relations and stationary state conditions into account, in addition to the auxiliary matrices $T_j$ and $Z_i$. The positivity matrix was $126 \times 126$ and the $A$, $B$, $C$ and auxiliary matrices were $120 \times 120$.}
	\begin{tabular}{|c|c|c|}
		\hline
		\textbf{Setup}& \textbf{Solver} & \textbf{Runtime}\\ \hline
		$g>0$, $L=8$& SDPA-DD & $\sim$ a few minutes  \\ \hline
		$g>0$, $L=10$& MOSEK & $\sim$ 10 seconds $-$ 1 minute \\ \hline
		$g<0$, $L=10$ infeasibility test& SDPA-DD&$\sim$ a few hours  \\ \hline
	\end{tabular}
\end{table}

As shown in the tables, the SDP solver time is relatively short. The main computing bottleneck in extending these results to longer length words is setting up the positivity matrices and reducing the number of variables using cyclicity of the trace and stationary state conditions. For example, at $L=10$ it takes approximately $\sim41$ minutes and $\sim46.5$ GB of memory to generate the symbolic SDP matrices.

\section{Fermions and long strings}\label{app:MQMrev}

In this Appendix, we brief review the fermion representation of the one-matrix QM, in the singlet as well as non-singlet sectors, largely following \cite{Marchesini:1979yq,Klebanov:1991qa,Marino_2015,Balthazar:2018qdv}, and derive an effective Hamiltonian for long string interactions that captures leading order effects in the low temperature expansion.

Here we consider the Hamiltonian (\ref{1matH}) without the traceless constraint on $X$ and $P$, as the traceless constraint is unimportant in the large $N$ limit. We begin by writing $X=\Omega^{-1}\Lambda\Omega$, where $\Lambda=\text{diag}\left(\lambda_1,\cdots,\lambda_N\right)$ is a diagonal matrix, and $\Omega$ is a unitary matrix whose elements are denoted $\Omega_{ia}$. While the indices $i$ and $a$ both range from 1 to $N$, they play different roles. In particular, the index $i$ labels the eigenvalues $\lambda_i$, whereas the index $a$ transforms in the fundamental representation of the $U(N)$ global symmetry.
The Hamiltonian can be equivalently expressed in terms of $\Lambda$ and $\Omega$ as
\ie
H=\sum_{i=1}^N\left[ -{1\over2}{\partial^2\over\partial \lambda_i^2}+V(\lambda_i)\right] +{1\over2}\sum_{i\neq j}\left[-{1\over\lambda_i-\lambda_j}{\partial\over\partial\lambda_i}+{R_{ij}R_{ji}\over(\lambda_i-\lambda_j)^2}\right],
\fe
where $R_{ij}\equiv\sum_a\Omega_{ia}{\partial\over\partial\Omega_{ja}}$. Note that while $R_{ij}$ obey the commutation relations of a $u(N)$ algebra, they do {\it not} transform under the $U(N)$ global symmetry. Introducing $Q=\prod_{i<j}(\lambda_i-\lambda_j)$, we can write $H=Q^{-1}H'Q$, where
\ie
H'=\sum_{i=1}^N\left[ -{1\over2}{\partial^2\over\partial \lambda_i^2}+V(\lambda_i)\right] +{1\over2}\sum_{i\neq j}{R_{ij}R_{ji}\over(\lambda_i-\lambda_j)^2} .
\fe
The Schr\"odinger equation $H\psi(\Lambda,\Omega) = E\psi(\Lambda,\Omega)$ is then equivalent to
\ie
H'\psi'(\Lambda,\Omega) = E\psi'(\Lambda,\Omega),~~~~\psi'(\Lambda,\Omega)\equiv Q\psi(\Lambda,\Omega).
\fe

\subsection{The singlet sector}\label{app:singlet}
Singlet wave function $\psi'_s$ does not depend on $\Omega$ and thus we have 
\ie\label{fermH}
H'\psi'_s(\Lambda)=\sum_{i=1}^N\left[ -{1\over2}{\partial^2\over\partial \lambda_i^2}+V(\lambda_i)\right] \psi'_s(\Lambda).
\fe
Action of $S_N$ gauge symmetry permuting the eigenvalues $\lambda_i$, when acting on $\psi'_s(\Lambda)$, produces extra minus signs due to the factor $Q$. Therefore, $\psi'_s(\Lambda)$ should be antisymmetric under the exchange of a pair of eigenvalues and the problem reduces to non-interacting fermions each of which are subject to the potential $V(\lambda)$. The ground state is simply described by $N$ fermions filling in the lowest $N$ energy eigenstates.

In 't Hooft large $N$ limit, semiclassical approximation can be employed to compute the energy and eigenvalue distribution of the ground state. Introducing the rescaled potential $v(\lambda)$ such that $V(\lambda)=Nv(\lambda/\sqrt{N})$, the semiclassically quantized phase-space area is given by
\ie
NJ(e)={N\over\pi}\int_{\lambda_-(e)}^{\lambda_+(e)}d\lambda\sqrt{2(e-v(\lambda))},
\fe
where $Ne$ is the energy of a single fermion subject to the potential $V(\lambda)$ and $\lambda_{\pm}(e)$ are turning points of $v(\lambda)$. Fermi energy $e_f$ is defined as the solution to $J(e_f)=1$. The ground state energy at the leading order in large $N$ is then given by
\ie
E_0=e_0N^2,~~e_0=e_f-{1\over3\pi}\int_{\lambda_-(e_f)}^{\lambda_+(e_f)}d\lambda\left(2(e_f-v(\lambda))\right)^{3\over2},
\fe
and the ground state eigenvalue distribution is given by
\ie
\R(\lambda)={1\over\pi}\sqrt{2(e_f-v(\lambda))}.
\fe

At large $N$, unbounded potentials may still lead to singlet ground states of finite energies. We can for example take $V(\lambda)={\lambda^2\over2}+{g\over N}\lambda^4$ with negative $g$. There is a potential well around $\lambda\sim0$ which can accomodate fermion energy levels, and there is no tunneling to the outside of the well in the 't Hooft limit. Therefore, as long as the Fermi level $e_f$ is smaller than the maximum of the potential $v(\lambda)={\lambda^2\over2}+g\lambda^4$, states obtained by filling in the energy levels within the well are stable and carry finite energies. Once $g$ is negative enough such that $e_f$ is greater than the maximum of the potential, there is no finite energy ground state. This transition happens at the critical coupling
\ie
g=g_c\equiv -{\sqrt{2}\over6\pi}.
\fe

\subsection{The adjoint sector}\label{app:singleadj}

In the adjoint sector of the MQM, wave functions a priori take the form
$\Omega^{\dagger}_{bj}\Omega_{ia} f_{ij}(\Lambda)$. A $U(1)^N$ gauge redundancy of the $(\Lambda, \Omega)$ parameterization further leads to the restriction $i=j$,\footnote{This is referred to as the zero weight condition in \cite{Gross:1990md,Maldacena:2005hi,Balthazar:2018qdv}.} and moreover the wave function should be invariant under the $S_N$ permutation on the index $i$ of $\Omega_{ia}$ along with the eigenvalue $\lambda_i$'s.
We will denote by $|ia; jb\rangle$ the state that corresponds to the basis wave function $\Omega^\dagger_{bj}\Omega_{ia}$,\footnote{Strictly speaking, we should subtract off ${1\over N} \delta_{ab} \delta_{ij}$, but this will be unimportant in the large $N$ limit of interest.}  normalized such that $\langle kc;\ell d | ia; jb\rangle=\delta_{ac}\delta_{ik}\delta_{bd}\delta_{j\ell}$. An ansatz for the energy eigenstate in the adjoint sector is
\ie
|w\rangle_{ab}=\sum_{i=1}^N w(\lambda_i)\psi_0(\lambda_1,\cdots,\lambda_N)|ia;ib\rangle,
\fe
where $\psi_0(\lambda_1,\cdots,\lambda_N)$ is the unit-normalized ground state wave function in the singlet sector. The spectral problem becomes
\ie
H'|w\rangle_{ab}&=\sum_{i=1}^N\left[ -{1\over2}\partial_{\lambda_i}^2 w(\lambda_i)-\partial_{\lambda_i}w(\lambda_i)\partial_{\lambda_i}+E_0w(\lambda_i)+\sum_{j\neq i}{w(\lambda_i)-w(\lambda_j)\over(\lambda_i-\lambda_j)^2}\right]\psi_0(\lambda_1,\cdots,\lambda_N)|ia;ib\rangle
\\
&=E|w\rangle_{ab}.
\fe
At large $N$, the last term in the bracket dominates \cite{Marchesini:1979yq}. The energy differences $\Delta_n$ ($n=1,2,\cdots$) against the singlet ground state is thus obtained by the following singular integral eigenvalue equation (with $\Delta_n$ being eigenvalues and $w_n(\lambda)$ being eigenfunctions)
\ie
\int_{\lambda_-(e)}^{\lambda_+(e)}d\lambda'\R(\lambda'){w_n(\lambda)-w_n(\lambda')\over(\lambda-\lambda')^2}=\Delta_nw_n(\lambda),
\fe
with orthogonality $\int_{\lambda_-(e)}^{\lambda_+(e)}d\lambda\R(\lambda)w_n(\lambda)=0$ and normalization $\int_{\lambda_-(e)}^{\lambda_+(e)}d\lambda\R(\lambda)|w_n(\lambda)|^2=1$ conditions. Note that each eigenstate has $N^2$ degeneracy since $H'$ does not act nontrivially on $a,b$ indices. The following results were computed in \cite{Marchesini:1979yq} for $v(\lambda)={\lambda^2\over2}+g\lambda^4$ ($\Delta_1$ is the smallest of $\Delta_n$'s)
\ie\label{SingAdjE}
g&=2~&:~~e_0=0.8654577,~~\Delta_1 =2.1281936
\\
g&=50~&:~~e_0=2.2167524,~~\Delta_1 =5.7594935
\fe

\subsection{Multi-adjoint sectors}\label{app:adjint}

The spectrum of multi-adjoint sectors can be understood as that of multiple long strings, each occupying a state in the adjoint sector. In the low temperature limit, the long strings effectively weakly interact with one another. The leading interaction between a pair of long strings can be determined by inspecting the matrix elements of the Hamiltonian in the bi-adjoint sector. We begin with the basis bi-adjoint states
\ie
|w\rangle_{(n,ab),(m,cd)}=\sum_{i,j=1}^N w_n(\lambda_i)w_m(\lambda_j)\psi_0(\lambda_1,\cdots,\lambda_N) |i a, jc;ib,jd\rangle,
\fe
where $|i a, jc;{i'}b,{j'}d\rangle$ corresponds to the wave function $\Omega^\dagger_{bi'}\Omega_{ia}\Omega^\dagger_{dj'}\Omega_{jc}$. Note that the $U(1)^N$ gauge redundancy requires $(i,j) = (i',j')$ or $(j',i')$. A straightforward computation produces the following action of the interaction Hamiltonian \cite{Balthazar:2018qdv}
\ie
{}&\left[ {1\over2}\sum_{i\neq j}{R_{ij}R_{ji}\over(\lambda_i-\lambda_j)^2}-(\Delta_n+\Delta_m)\right] |w\rangle_{(n,ab),(m,cd)}
\\
&=-\sum_{i\neq j}{\psi_0(\lambda_1,\cdots,\lambda_N)\over(\lambda_i-\lambda_j)^2}(w_n(\lambda_i)-w_n(\lambda_j))(w_m(\lambda_i)-w_m(\lambda_j)) |i a, jc;jb,id\rangle. 
\fe
Therefore, the matrix element of the interaction Hamiltonian between a pair of bi-adjoint states is given by
\ie\label{eqn:longint}
{}_{(k,ef),(\ell,gh)}\langle w| \left[ {1\over2}\sum_{i\neq j}{R_{ij}R_{ji}\over(\lambda_i-\lambda_j)^2}-(\Delta_n+\Delta_m)\right] |w\rangle_{(n,ab),(m,cd)}={h_{nmk\ell}\over N}(\delta_{ae}\delta_{bh}\delta_{cg}\delta_{df}+\delta_{ag}\delta_{bf}\delta_{ce}\delta_{dh}),
\fe
where the long string coupling coefficients $h_{nmk\ell}$ are given by
\ie\label{hcoeff}
h_{nmk\ell}=-\int_{\lambda_-(e)}^{\lambda_+(e)} d\lambda d\lambda' \rho(\lambda)\rho(\lambda'){w_k^*(\lambda)w_\ell^*(\lambda')(w_n(\lambda)-w_n(\lambda'))(w_m(\lambda)-w_m(\lambda')) \over(\lambda-\lambda')^2 }.
\fe
Upon introducing the creation and annihilation operators $\left(\mathfrak{a}^\dagger\right)_{n,ab}$ and $\left(\mathfrak{a}\right)_{n,ba}$ for a single adjoint energy eigenstate of energy $\Delta_n$ obeying $\left[\left(\mathfrak{a}\right)_{n,ab}, \left(\mathfrak{a}^\dagger\right)_{m,cd}\right]=\delta_{nm}\delta_{ad}\delta_{bc}$, it is straightforward to derive
\ie
{}&\langle0|\left(\mathfrak{a}\right)_{n_3,fe}\left(\mathfrak{a}\right)_{n_4,hg}\left(\sum_{c_1,c_2,c_3,c_4=1}^N \left(\mathfrak{a}^\dagger\right)_{n_5,c_1c_2}\left(\mathfrak{a}^\dagger\right)_{n_6,c_3c_4}\left(\mathfrak{a}\right)_{n_7,c_4c_1}\left(\mathfrak{a}\right)_{n_8,c_2c_3}\right)\left(\mathfrak{a}^\dagger\right)_{n_1,ab}\left(\mathfrak{a}^\dagger\right)_{n_2,cd}|0\rangle
\\
&=(\delta_{n_1n_7}\delta_{n_2n_8}\delta_{n_3n_6}\delta_{n_4n_5}+\delta_{n_1n_8}\delta_{n_2n_7}\delta_{n_3n_5}\delta_{n_4n_6})\delta_{ag}\delta_{bf}\delta_{ce}\delta_{dh}
\\
&~~~+(\delta_{n_1n_7}\delta_{n_2n_8}\delta_{n_3n_5}\delta_{n_4n_6}+\delta_{n_1n_8}\delta_{n_2n_7}\delta_{n_3n_6}\delta_{n_4n_5})\delta_{ae}\delta_{bh}\delta_{cg}\delta_{df}.
\fe
Using $h_{nmk\ell}=h_{mnk\ell}=h_{nm\ell k}$, the multi-particle Hamiltonian that captures the bi-adjoint pairwise interaction (\ref{eqn:longint}) is therefore given by (\ref{lsham}).

\newpage
\bibliography{refs}
\bibliographystyle{JHEP}
\end{document}